\documentclass[12pt, fleqn]{extarticle}

\usepackage[utf8]{inputenc}
\usepackage{amsmath, amssymb, mathtools, mathrsfs, gensymb, physics, bm, upgreek}
\usepackage[low-sup]{subdepth}
\usepackage[shortlabels]{enumitem}
\usepackage[toc]{appendix}
\usepackage{graphicx}
\usepackage[raggedright]{sidecap}
\usepackage[labelfont=bf]{caption}
\usepackage{setspace}

\usepackage{hyperref}
\hypersetup{colorlinks=true, urlcolor=blue, linkcolor=blue, citecolor=blue}

\usepackage[letterpaper, left=1.2in, right=1.2in, top=1in, bottom=1in]{geometry}
\usepackage{titlesec}
\titleformat{\section}{\large\bfseries}{\thesection.}{10pt}{\large}
\titleformat{\subsection}{\normalsize\bfseries}{\thesubsection.}{10pt}{\normalsize}

\newcommand{\eq}[1]{\begin{equation} #1 \end{equation} }
\newcommand{\eqa}[1]{\begin{align*} #1 \numberthis \end{align*}}

\newcommand{\numberthis}{\addtocounter{equation}{1}\tag{\theequation}}

\newcommand{\p}{\partial}
\newcommand{\cov}{\nabla}

\newcommand{\A}{\mu}  
\newcommand{\B}{\nu}

\renewcommand{\expval}[1]{\langle #1 \rangle}

\begin{document}

\begin{center}
\textbf{\large Evolving Dark Energy from the Back-Reaction of Cosmological Perturbations}

\vspace{10pt}

\normalsize Vincent Comeau\footnote{Email: \href{mailto:vincent.comeau@mail.mcgill.ca}{vincent.comeau@mail.mcgill.ca}} \\[5pt]

\textit{Department of Physics -- McGill University} \\
(5 June 2026)
\vspace{5pt}

\end{center}

\begin{abstract}
\singlespace
We argue that the back-reaction of inhomogeneities, when combined with a pure cosmological constant, produces an effective dark energy whose equation of state can evolve with time. Our analytical computation agrees with recent numerical simulations, which have shown that the back-reaction of inhomogeneities can impact dark energy around the present time. Here, its equation of state is derived from an average version of the Friedmann equations, for which the average is taken over a fixed volume of matter particles (not necessarily global). When treating the inhomogeneities as small perturbations around a homogeneous background, dark energy acquires a time dependence which reduces to a boundary term, at least at quadratic order in the scalar modes of the perturbations. Thus, it becomes negligible for a global average. On the other hand, when treating instead the spatial derivatives of the inhomogeneities perturbatively in a gradient expansion, dark energy gets the same time dependence, which no longer reduces to a boundary term. This time dependence can be significant even for a global average or when the constant spatial curvature induced by the inhomogeneities is negligible. It also keeps the same sign around the present time, thus causing dark energy to remain either always phantom or never phantom, not allowing for any crossing from one type of behaviour to the other. 
\vspace{10pt}
\end{abstract}

\normalsize

\section{Introduction and summary}

\subsection{Current state of the question}

According to the standard cosmological model, an important component of the universe today takes the form of ``dark energy''. A key research objective has been to determine if dark energy is just a cosmological constant, or if it corresponds to something else whose energy density and pressure evolves with the expansion of the universe. To constrain the equation of state of dark energy, many models have been proposed \cite{Sahni:2006, Li:2011}. For example, in the commonly used Chevallier-Polarski-Linder parametrization \cite{Chevallier:2000, Linder:2002}, the equation of state of dark energy is expanded in terms of the scale factor around the present time,
\eq{ \label{CPL parametrization}
w_\text{DE} = -1 + \delta w_0 + \delta w_1(1-a) + \order{(1-a)^2}\,,
}
where $a$ represents the scale factor, which is set to reach $a=1$ at the present time. Here, $\delta w_0$ and $\delta w_1$ are two parameters that can be constrained by observations. These parameters vanish for a pure cosmological constant, but not for other models of dark energy. The equation of state could also be expanded in terms of other quantities, such as the redshift \cite{Cooray:1999} or the logarithm of the scale factor \cite{Gerke:2002}. Although these alternative parametrizations will not be considered in this paper, their parameters could be derived from our results.

A lot of work has been done in recent years to try to explain the origin of dark energy in terms of various mechanisms. For example,  many have proposed that dark energy could be an exotic form of matter not yet detected, represented by additional fields coupled with gravity as in quintessence \cite{Li:2011}. Alternatively, some have argued that dark energy could simply be the result of the nonlinear effects of inhomogeneities, when local quantities are averaged over a certain volume \cite{Brandenberger:2002, Rasanen:2003, Ellis:2005, Buchert:2007, Buchert:2011, Buchert:2015, Ellis:2011, Clarkson:2011, Kolb:2011}. Others have argued that these back-reaction effects should remain negligible \cite{Geshnizjani:2005, Green:2011, Green:2015}. Although the question is still open, recent numerical computations using the full equations of general relativity seem to indicate that the back-reaction of inhomogeneities is too small to account completely for dark energy \cite{Adamek:2017, Macpherson:2018, Oestreicher:2024}. 

Hence, in this paper, we do not try to explain the presence of a cosmological constant from the back-reaction of inhomogeneities (or any other mechanism). We assume that a cosmological constant is present in the universe, in addition to cold matter. We neglect the other components of the universe usually included in the standard model, specifically its radiation and constant spatial curvature, as their energy densities become negligible around present time.

Our goal is more modest. Recent numerical computations have shown that evolving dark energy can emerge from a pure cosmological constant combined with the back-reaction effects of inhomogeneities \cite{Ginat:2026, Macpherson:2026, Alvarez:2025}. In this paper, we will reach the same conclusion analytically, through a perturbative calculation which has never been done before, so far as we are aware. We will show that inhomogeneities can yield a non-zero value for the parameters $\delta w_0$ and $\delta w_1$ in the equation of state of dark energy, even when the universe contains nothing more than cold matter and a pure cosmological constant. At the very least, the back-reaction of inhomogeneities can affect the value of the parameters in ways which should be taken into account. Back-reaction, if it does not completely explain dark energy, does at least contribute to its evolving nature.

\subsection{Approach followed in this paper}

In this paper, we will follow closely the approach developed by T. Buchert and his collaborators \cite{Buchert:2000, Buchert:2001, Buchert:2019}. We will consider a spacetime containing cold matter with negligible pressure, and a cosmological constant. When the matter is not uniformly distributed, as in our actual universe, the local expansion rate differs from point to point in space. We will compute the average value of the expansion rate over a spatial volume occupied by a fixed set of matter particles, located at the same proper time along their trajectories. Choosing a fixed set of particles as the averaging volume ensures that the average is physically meaningful.

We will then write an ``average version'' of the Friedmann equations, involving the expansion rate and matter density averaged over the chosen set of particles. These average Friedmann equations will involve additional terms induced by the inhomogeneities. Some of these terms behave like constant spatial curvature, with an energy density $\rho_K$ and a pressure $p_K = -\frac13 \rho_K$ decaying with the square of the scale factor. The other terms will be grouped as a ``back-reaction'' effective fluid, with an energy density $\rho_\text{br}$ and a pressure $p_\text{br}$.

The combination of these back-reaction terms with the cosmological constant present in the spacetime can be defined as the effective dark energy, having an energy density and a pressure given by
\eqa{
\rho_\text{DE} &= \frac{\Lambda}{\kappa} + \rho_\text{br}\,, \\[5pt]
p_\text{DE} &= -\frac{\Lambda}{\kappa} + p_\text{br}\,,
}
where $\Lambda$ is the cosmological constant, $\kappa=8\pi G_N$, and $G_N$ is Newton's gravitational constant. This effective dark energy has an equation of state which is not necessarily constant. At leading order in the back-reaction terms, it is given by
\eq{ \label{DE-equation-state}
w_\text{DE} = \frac{p_\text{DE}}{\rho_\text{DE}} \,\approx\, -1 + \frac{\kappa(\rho_\text{br} + p_\text{br})}{\Lambda}\,.
}

Expanding around the present time, we can relate the back-reaction of inhomogeneities to the parameters $\delta w_0$ and $\delta w_1$ involved in the Chevallier-Polarski-Linder parametrization of dark energy.

\subsection{Equation of state from linear perturbations}

Before presenting our computation in more details, we will first summarize and analyze our findings, for the convenience of the reader. In section \ref{section-framework} of this paper, we explain our approach and notation. In section \ref{section-linear}, we treat the matter inhomogeneities as small perturbations around a homogenous background, following the standard theory of cosmological perturbations. We keep terms up to the quadratic order in the perturbations and neglect higher-order terms. At quadratic order, the scalar modes of the perturbations generate the following equation of state for dark energy, 
\eq{ \label{linear-intro}
w_\text{DE} = -1 - \frac{1}{4H_0^4}\left( \big\langle \p_i \big( \p_j A \,\p_i\p_j A\big)\big\rangle - \big\langle \p_i \big( \p_i A\,\cov^2 A\big)\big\rangle + \frac23 \expval{\cov^2 A}^2 \right) \delta \bar{w}(a)\,,
}
where $A(\bm{x})$ represents the amplitude of the scalar perturbations, and $H_0$ denotes the background expansion rate at the present time. Moreover, $\delta\bar{w}(a)$ is a function of the scale factor, which also depends on the density parameters for cold matter and the cosmological constant. It can be inferred from (\ref{linear-DE-equation}). When expanded around the present time, the corresponding parameters $\delta w_0$ and $\delta w_1$ of the Chevallier-Polarski-Linder model are then given by (\ref{linear-parameters}).

This equation of state is similar to the one obtained in \cite{Li:2007}, for a spacetime containing only cold matter, but no cosmological constant. In both cases, the back-reaction from the linear scalar modes at quadratic order reduces to a boundary term, although with a different time dependence. Hence, it becomes negligible when the average is taken globally, over all the matter particles. The scalar modes generate non-negligible back-reaction even for a global average only when their contribution beyond quadratic order is taken into account.

On the other hand, the tensor modes of the perturbations generate quadratic effects which do not vanish even for a global average. However, we have not (yet) been able to solve the wave equation satisfied by the linear tensor modes for a background containing both cold matter and a cosmological constant. When solving it perturbatively with a gradient expansion, we find an approximate solution which is formally similar to the scalar modes, involving the same dependence on the scale factor, but which does not reduce to a boundary term.

\subsection{Equation of state from a gradient expansion}

For these reasons, a different approach is needed. In section \ref{section-gradient} of this paper, we no longer treat the inhomogeneities as small perturbations, but treat instead their spatial derivatives perturbatively, following a gradient expansion. In that case, the spatial components of the metric are proportional to a constant inhomogeneous background $\bar{g}_{ij}(\bm{x})$, plus small perturbations which vary with time.

The first two orders of the gradient expansion allow us to generalize the previous results obtained from a homogeneous background. More specifically, the back-reaction terms involve the same time dependence as in the standard perturbative approach, but do not reduce to boundary terms and thus remain significant even when the average is taken globally. At the second order in the gradient expansion, the equation of state of dark energy is indeed
\eq{ \label{gradient-intro}
w_\text{DE} = -1 - \frac{1}{H_0^4}\left(\expval{\bar{R}_i^j\,\bar{R}_j^i} - \frac38 \expval{\bar{R}^2} + \frac{1}{24} \expval{\bar{R}}^2 \right) \delta\bar{w}(a)\,,
}
where $\delta \bar{w}(a)$ denotes the same function of the scale factor as in (\ref{linear-intro}). Moreover, $\bar{R}_{ij}$ represents the Ricci tensor for the three-dimensional background metric $\bar{g}_{ij}(\bm{x})$, and $\bar{R} = \bar{g}^{ij}\,\bar{R}_{ij}$ is the corresponding Ricci scalar.

In particular, if the inhomogeneities are treated once again as small perturbations, with $\bar{g}_{ij}(\bm{x}) = \delta_{ij}+\delta_{ij}\,A(\bm{x})$, this equation of state then coincides with the one obtained from the standard perturbative approach. In other words, (\ref{gradient-intro}) reduces to (\ref{linear-intro}) at quadratic order in $A(\bm{x})$. Hence, the gradient expansion allows us to compute the back-reaction of the scalar mode $A(\bm{x})$ not just at quadratic order, but at all orders in its amplitude. It also accounts for the vector and tensor modes induced by this scalar mode, as well as its interactions with other modes.

At the first order in the gradient expansion, the background inhomogeneities do not produce any back-reaction, but generate instead constant spatial curvature. This confirms the findings of \cite{Geshnizjani:2005, Blachier:2023}, which show that only constant spatial curvature is generated by the inhomogeneities in the long-wavelength limit. Our computation goes one step further, including terms at the second order in the gradient expansion, with up to four spatial derivatives of the background metric, which is why we find significant back-reaction.

Incidentally, the effect of inhomogeneities on dark energy can remain significant even when their contribution to the constant spatial curvature is not. At the leading order in the gradient expansion, the energy density of the constant spatial curvature is proportional to the average Ricci scalar of the inhomogeneous background, $\rho_K \sim \expval{\bar{R}}$. In the case when this energy density is negligible, $\rho_K \approx 0$, the same also holds for the average Ricci scalar, $\expval{\bar{R}} \approx 0$. However, it does not follow that its variance or the average of the Ricci tensor squared should also be negligible. Hence, inhomogeneities can still have an impact on dark energy even in that case.

\begin{SCfigure}[0.6][h]
\includegraphics[width=8cm]{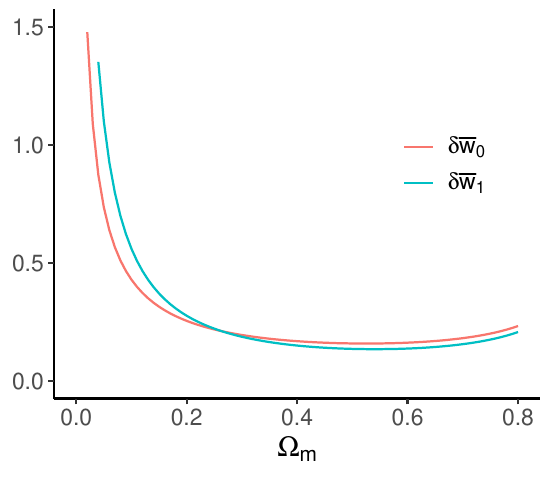}
\caption{Parameters $\delta \bar{w}_0$ and $\delta \bar{w}_1$ for different values of the density parameter for cold matter at the present time. \vspace{20pt}} 
\label{figure-parameters}
\end{SCfigure}

The previous equation of state is valid at all times. When expanded around the present time, it takes the form
\eq{
w_\text{DE} = -1 - \frac{1}{H_0^4}\left(\expval{\bar{R}_i^j\,\bar{R}_j^i} - \frac38 \expval{\bar{R}^2} + \frac{1}{24} \expval{\bar{R}}^2 \right) \Big(\delta\bar{w}_0 + \delta\bar{w}_1(1-a)\Big)\,,
}
where $\delta\bar{w}_0$ and $\delta\bar{w}_1$ are two parameters which can be inferred from (\ref{gradient-parameters}) and are plotted in figure \ref{figure-parameters}. They depend exclusively on the density parameters of cold matter and the cosmological constant. For example, when $\Omega_m = 0.3$ and $\Omega_\Lambda = 0.7$, we find $\delta\bar{w}_0 = 0.1946$ and $\delta\bar{w}_1 = 0.1871$. 

The sign of the parameters $\delta w_0$ and $\delta w_1$ depends on the spatial curvature terms, which could be either positive or negative, as well as on the parameters $\delta \bar{w}_0$ and $\delta \bar{w}_1$. As shown in figure \ref{figure-parameters}, these parameters are very similar to each other and remain positive for all values of the density parameter for cold matter. Hence, $\delta w_0$ and $\delta w_1$ should also be very similar to each other and should bear the same sign, being either both positive or both negative. 

\begin{SCfigure}[0.6][h]
\includegraphics[width=8cm]{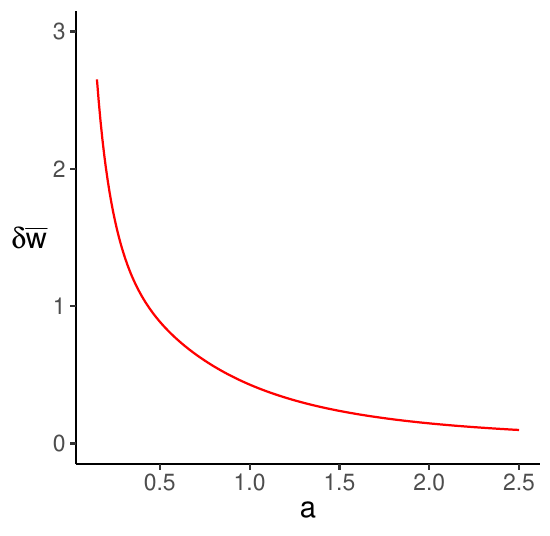}
\caption{Values of $\delta \bar{w}$ as a function of the scale factor, for density parameters given by $\Omega_m = 0.3$ and $\Omega_\Lambda = 0.7$. \vspace{20pt}} 
\label{figure-function}
\end{SCfigure}

Therefore, the effective dark energy generated by the cosmological constant and the inhomogeneities either remains phantom, with $w_\text{DE} < -1$ at all times, or is never phantom, with $w_\text{DE} > -1$. No transition from one type of behaviour to the other is possible around the present time. This conclusion is confirmed when the full time dependence of the equation of state is taken into account. For example, figure \ref{figure-function} shows that the function $\delta\bar{w}(a)$ remains positive all the way from early to late times, whenever the scale factor becomes large enough compared to its initial value. Dark energy is thus either always phantom or never so, depending on the sign of the spatial curvature terms. Incidentally, the figure also shows that the time dependence of dark energy becomes negligible far into the future, when the spacetime is dominated by the cosmological constant.  

This feature of our result differs from the numerical analysis conducted in \cite{Ginat:2026}, which finds that the two parameters of the Chevallier-Polarski-Linder model usually have opposite signs, with $\delta w_0$ being positive and $\delta w_1$ negative. Current observational constraints also seem to favour the parameters having different signs, thus possibly allowing a transition between phantom and non-phantom behaviours \cite{DES:2024, DESI:2024, RoyChoudhury:2024, Park:2024}. However, as argued in \cite{Shlivko:2024, Payeur:2024}, alternative parametrizations of dark energy can alleviate or significantly modify these observational constraints, eliminating in particular the evidence for a phantom crossing.

Further work is needed to fully explain the difference between our results and that of \cite{Ginat:2026}, and more generally to determine how our approach can be compared with numerical simulations. In any case, both types of computations seem to show that the back-reaction of inhomogeneities can meaningfully impact the equation of state of dark energy and contribute to its evolving nature.

\subsection{Limits and future work}

The analysis conducted in this paper could be extended in various ways. For example, it would be straightforward to expand the equation of state of dark energy around present time beyond the linear order, using its full time-dependence as provided in this paper. Future observations might become precise enough to constrain not just the two first parameters in the expansion, $\delta w_0$ and $\delta w_1$, but higher-order parameters, which could easily be obtained from the results of this paper.

Furthermore, our findings are valid up to certain perturbative orders. On the one hand, it would be possible to extend our computation from section \ref{section-gradient} beyond the second order in the gradient expansion, to determine in particular if a phantom crossing can take place when higher-order terms are included. This will be examined in future work. On the other hand, it would be a lot more challenging to extend our computation from section \ref{section-linear} beyond the quadratic order in the perturbations, since the equations involved at higher orders cannot be solved so easily. We could not even solve fully the wave equation satisfied by the linear tensor modes, let alone the same wave equation in the presence of a source term with non-trivial time dependence.

\section{General framework} \label{section-framework}

\subsection{Average and averaging volume}

In this paper, we will compute averages over the volume occupied by a fixed set of matter particles, located at the same proper time along their trajectories, assuming that their proper time coincides with the background time initially. The average will not necessarily be a global average, taken over all the particles contained in the spacetime. For this reason, we will not  neglect boundary terms nor impose periodic boundary conditions. Similarly, we will not assume that the averaging volume is homogeneous on average, or that perturbations have an average of zero. This might be true for the whole spacetime, but not necessarily for the set of particles over which the average will be computed.

Furthermore, since the average is not taken globally, it should not be taken over a fixed region of spatial coordinates. As argued in \cite{Buchert:2019}, matter would then constantly flow in and out of that region, which could lead to unphysical back-reaction effects. Generally speaking, spatial coordinates are just labels and do not bear any physical meaning. On the other hand, averaging over a fixed set of particles is physically well-defined and ensures that the mass of matter contained in the averaging volume is conserved.  

Let $g_{ij}$ represent the three-dimensional metric of the volume containing the chosen set of matter particles, at a given proper time, after the scale factor of the background has been factored out. For general coordinates, this metric would be different from the spatial components of the spacetime metric, although they will coincide in this paper because of our choice of coordinates. Following \cite{Buchert:2000}, the average of a function $F$ over the volume of particles at a given proper time is
\eq{ \label{definition-average}
\expval{F} = \frac{1}{V} \int \dd[3]{x} \sqrt{g^{(3)}}\,a^3\,F(t, \bm{x})\,,
}
where $a(t)$ is the scale factor of the background, $g^{(3)}$ is the determinant of the spatial metric $g_{ij}$, and $V$ represents the volume occupied by the chosen particles, 
\eq{ \label{volume}
V = \int \dd[3]{x}\,\sqrt{g^{(3)}}\,a^3 \,.
}
This volume can be used to define an effective scale factor $\bar{a}$, whose cube is proportional to the volume occupied by the particles at a given time,
\eq{ \label{effective-a}
V = C\,\bar{a}^3\,,
}
where the proportionality constant $C$ can be fixed, for example, by the initial value of the volume or its value at the present time.

\subsection{Choice of Lagrangian coordinates}

For general spacetime coordinates, matter particles do not remain at the same spatial coordinates at all times. Denoting their trajectory as $x^\A(T)$, where $T$ represents their proper time, the four-velocity of the particles is then
\eq{
u^\A = \dv{x^\A}{T}\,.
}

For a spacetime metric $g_{\A\B}$ whose time component is negative, the four-velocity of the particles has a norm of $g_{\A\B}\,u^\A \,u^\B = -1$. Since we want to compute averages over a volume of particles at the same proper time, it is convenient to choose so-called ``Lagrangian coordinates'' associated with the matter flow. In that case, matter particles remain at the same spatial coordinates at all times, 
\eq{
u^i = \dv{x^i}{T} = 0\,,
}
and the time coordinate coincides with the proper time of the particles, $T=t$,
\eq{
u^0 = \dv{t}{T} = 1\,.
}

The proper time is fully determined up to the choice of its initial value, which can vary in space, $T = t + T_\text{in}(\bm{x})$. Assuming that the proper time of the particles initially coincides with the background time, we can take $T=t$. An average over a fixed set of particles can then be taken over a fixed region of spatial coordinates $\bm{x}$, at a given coordinate time $t$. Furthermore, in Lagrangian coordinates, the norm of the four-velocity implies that $g_{00} = -1$. Similarly, cold matter particles with negligible pressure move along geodesics, and their four-velocity satisfies $u^\B \cov_\B u^\A = 0$, which implies that $\dot{g}_{0i} = 0$, where a dot represents a derivative with respect to time. The integration constant corresponds to a residual gauge freedom and can thus be set to zero, $g_{0i} = 0$. Hence, in Lagrangian coordinates, the spacetime metric can be written in the synchronous-comoving gauge, that is,
\eq{ \label{metric}
\dd{s}^2 = -\dd{t}^2 + a(t)^2\,g_{ij}(t,\bm{x})\,\dd{x}^i\,\dd{x}^j\,,
}
where $a(t)$ denotes the scale factor of the background. For convenience, $g_{ij}$ represents the spatial components of the metric after the scale factor has been factored out.

\subsection{Background}

In section \ref{section-linear} of this paper, we will treat inhomogeneities as small perturbations around a homogeneous background containing cold matter and a cosmological constant. In that background, matter is uniformly distributed and has an energy density $\bar{\rho}_m$, which evolves with the scale factor as
\eq{
\bar{\rho}_m = \frac{\rho_{m\,0}}{a^3}\,,
}
where $\rho_{m\,0}$ represents the value of the background matter density at present time, when the scale factor reaches $a=1$. Moreover, according to recent cosmological observations \cite{Planck:2018}, the universe has little spatial curvature on large scales and can thus be described approximately as a spatially flat background, with $g_{ij} = \delta_{ij}$. In that case, using the first Friedmann equation, the background expansion rate can be written as
\eq{ \label{background-H}
H = \frac{\,\dot{a}\,}{a} = H_0\sqrt{\frac{\Omega_m}{a^3} + \Omega_\Lambda}\,,
}
where $H_0$ denotes its value at present time. In this paper, a dot indicates a derivative with respect to time. Here, $\Omega_m$ and $\Omega_\Lambda$ are the density parameters for cold matter and the cosmological constant,
\eqa{
\Omega_m &= \frac{\kappa \rho_{m\,0}}{3H_0^2}\,, \\[5pt]
\Omega_\Lambda &= \frac{\Lambda}{3H_0^2}\,,
}
with $\Omega_m + \Omega_\Lambda = 1$ in the absence of any other matter component.

\subsection{Average expansion rate}

Since the universe is not actually homogeneous, the uneven matter distribution affects the value of its expansion rate. Following \cite{Ehlers:1993}, the local expansion rate can be defined in terms of the four-velocity of the matter particles moving in the spacetime. In Lagrangian coordinates, we find
\eq{
\Theta = \frac13 \,\cov_\A u^\A\, = H + \frac16\, L\,,
}
where $L=g^{ij}\,\dot{g}_{ij}$. As opposed to the standard definition, we choose to add a factor of $\frac13$ to ensure that the expansion rate reduces to its background value in the absence of inhomogeneities. Equivalently, the average expansion rate over the chosen set of particles can also be defined in terms of how their volume changes with time, as the spacetime expands and particles move further apart. Taking the time derivative of the volume occupied by the particles (\ref{volume}), we get
\eq{
\expval{\Theta} = \frac{\dot{V}}{3V} = H + \frac16 \,\expval{L}\,,
}
which confirms that the two definitions are equivalent. Finally, as shown in \cite{Buchert:2000}, the local expansion rate appears in a ``commutation rule'' useful when taking the time derivative of an average. In Lagrangian coordinates, it takes the form
\eq{ \label{commutation-rule}
\dv{}{t}\expval{F} = \expval{\dot{F}} + 3\,\expval{F\,\Theta} - 3\,\expval{F}\expval{\Theta}\,.
}

\subsection{Einstein equations in Lagrangian coordinates} \label{section-Einstein-equations}

We assume that the dynamics of spacetime are governed by general relativity. As an important difference with the work of T. Buchert and collaborators in \cite{Buchert:2000, Buchert:2019}, we will write and solve the full Einstein equations, either following the standard perturbative approach in section \ref{section-linear}, or with a gradient expansion in section \ref{section-gradient}. In Lagrangian coordinates, for a metric described by (\ref{metric}), the energy constraint of the Einstein equations takes the form
\eqa{ \label{Einstein-energy}
\kappa \rho_m + \Lambda = 3H^2 + HL + \frac18 L^2 - \frac18 L_i^j L_j^i + \frac{R}{2a^2}\,,
}
where $L_i^j$ is proportional to the extrinsic curvature of the three-dimensional space with metric $g_{ij}$, which in this case also corresponds to its time derivative,
\eqa{
&L_i^j = g^{jk}\,\dot{g}_{ik}\,, \\[5pt]
&L = L_i^i = g^{ij}\,\dot{g}_{ij}\,.
}
Similarly, in Lagrangian coordinates, the momentum constraint reads
\eq{ \label{Einstein-momentum}
\cov_j L_i^j - \cov_i L = 0\,,
}
where $\cov_i$ denotes the covariant derivative for the three-dimensional metric $g_{ij}$. Finally, the Einstein equations also imply the following evolution equations,
\eqa{ \label{Einstein-evolution}
&\dot{L}_i^j + 3H L_i^j + \frac12 L L_i^j + \frac18\,\delta_i^j\Big(L_k^l L_l^k - L^2\Big) + \frac{2}{a^2}\left(R_i^j -\frac14 R\,\delta_i^j\right) = 0\,, \\[5pt]
&\dot{L} + 3HL + \frac18 L^2 + \frac38 L_i^j L_j^i + \frac{R}{2a^2} = 0\,.
}

The second equation is obtained by taking the trace of the first one. Here, $R_{ij}$ represents the Ricci tensor for the three-dimensional metric $g_{ij}$, and $R$ is the corresponding Ricci scalar, $R = g^{ij} R_{ij}$.

\subsection{Average Friedmann equations}

We will write an ``average version'' of the Friedmann equations, involving the local expansion rate and matter density averaged over the chosen set of particles. Although the background considered in this paper is spatially flat, the most general Friedmann equations should include constant spatial curvature, carrying an energy density $\rho_K$ and a pressure $p_K = -\frac13 \rho_K$, with
\eq{ \label{constant-curvature}
\kappa \rho_K = \frac{K}{\bar{a}^2}\,,
} 
where $K$ is a constant, and $\bar{a}$ is the effective scale factor for the volume occupied by the chosen particles (\ref{effective-a}). Some of the terms induced by the inhomogeneities in the average Friedmann equations do behave as constant spatial curvature. The other terms, having a different dependence on the scale factor, will be grouped together as an effective ``back-reaction'' fluid, with an energy density $\rho_\text{br}$ and a pressure $p_\text{br}$. The average Friedmann equations can then be written as 
\eqa{
\kappa \expval{\rho_m} + \kappa\rho_\text{br} + \kappa\rho_K + \Lambda &= 3\,\expval{\Theta}^2\,, \\[5pt]
\kappa p_\text{br} + \kappa p_K - \Lambda &= -3\,\expval{\Theta}^2 - 2\,\dv{}{t}\expval{\Theta}\,.
}
Combining these equations with the Einstein equations (\ref{Einstein-energy}) and (\ref{Einstein-evolution}), and using the commutation rule (\ref{commutation-rule}), we find
\eqa{ \label{br-density-pressure}
&\kappa(\rho_\text{br} + \rho_K) = Q - \frac{\expval{R}}{2a^2}\,, \\[5pt]
&\kappa(p_\text{br} + p_K) = Q + \frac{\expval{R}}{6a^2}\,,
}
where $Q$ represents the so-called ``kinematical'' back-reaction,
\eq{ \label{kinematical}
Q = \frac18 \expval{L_i^j L_j^i} - \frac18 \expval{L^2} + \frac{1}{12} \expval{L}^2\,.
}

\subsection{Initial conditions} \label{initial-conditions}

In this paper, we will fix initial conditions at some time in the early stages of matter domination, when the scale factor is equal to $a_\text{in}$ and is much smaller than its present value, $a_\text{in} \ll 1$. Since the back-reaction of inhomogeneities will be evaluated around present time, the actual value of $a_\text{in}$ will not have any incidence on our results.

We choose initial conditions ensuring that no back-reaction is initially produced by the inhomogeneities. More specifically, we ask that the back-reaction density and pressure both vanish initially, in the initial stages of matter domination,
\eq{
\rho_\text{br}(a_\text{in}) = p_\text{br}(a_\text{in}) = 0\,.
}

These initial conditions basically amount to neglecting the decaying mode of the perturbations, or equivalently requiring that their time derivative vanishes initially\footnote{This can be seen from (\ref{linear-E-general}) in the case of the linear perturbations. On the one hand, neglecting the decaying mode of the perturbations amounts to taking $B=0$. On the other hand, requiring that the back-reaction vanishes initially leads to $B \sim a_\text{in}^{\frac52}$. Hence, at late times, when the scale factor is much larger than its initial value, the effect of $B$ becomes negligible, thus proving that these initial conditions are equivalent to neglecting the decaying mode.}. In that case, the effective dark energy corresponds initially to a pure cosmological constant. Its equation of state diverges from that of a cosmological constant only afterwards, as the spacetime expands and inhomogeneities evolve,
\eq{
w_\text{DE}(a_\text{in}) = -1\,.
}

\section{Linear perturbations} \label{section-linear}

\subsection{Standard perturbative approach}

This section represents our first attempt at computing the back-reaction of inhomogeneities and their contribution to the equation of state of dark energy. To do so, we will treat the inhomogeneities as small perturbations affecting an homogeneous background, following the standard theory of cosmological perturbations \cite{Mukhanov:1990}. We will keep terms quadratic in the perturbations and neglect higher-order terms.

A lot of work has been done to assess the back-reaction of perturbations up to quadratic order in their amplitude. For example, their effect on the average expansion rate of a spacetime containing only cold matter was computed in \cite{Rasanen:2003, Kolb:2004}, using various gauges. A computation similar to ours was done in \cite{Li:2007}, once again for a spacetime containing only cold matter, but no cosmological constant. The matter density perturbation was computed in \cite{Silveira:1994, Vale:2001} at linear order, for the case when a cosmological constant is also present in the spacetime. 

Perturbations can be decomposed into scalar, vector, and tensor modes, which evolve independently from one another at linear order. The spatial component of the metric can then be written as
\eq{
g_{ij} = \delta_{ij} + \psi\,\delta_{ij} + \p_i\p_j E + \p_i E_j + \p_j E_i + H_{ij}\,,
}
where $\psi$ and $E$ represent the scalar modes, $E_i$ are the vector modes with $\p_i E_i = 0$, and $H_{ij}$ are the traceless and transverse tensor modes, with $H_{ii} = 0$ and $\p_i H_{ij} = 0$. At linear order, the vector modes decay with the spacetime expansion and can thus be safely neglected. This is no longer true at higher orders, due to the vector modes induced by the linear scalar and tensor modes. We would have to take this effect into account only if we were to compute the back-reaction beyond quadratic order in the perturbations.

\subsection{Linear scalar modes}

Let us first focus on the scalar modes at linear order. The linearized Einstein equations imply that the mode $\psi$ does not change with time,
\eq{ \label{linear-psi}
\psi = A(\bm{x})\,.
}
Similarly, the mode $E$ satisfies the equation
\eq{ \label{equation-E}
\ddot{E} + 3H\,\dot{E} - \frac{1}{a^2}\,\psi = 0\,.
}
Two integrals needed to solve this equation are computed in appendix \ref{appendix-integrals}. Integrating the equation once with respect to time, and using (\ref{first-integral}), we find
\eqa{
\dot{E} &= \frac{A(\bm{x})}{a^3}\int \dd{t} a \,+\, \frac{B(\bm{x})}{a^3}\\[5pt]
&= \frac{2\,A(\bm{x})}{5H_0\sqrt{\Omega_m}\,a^{\frac12}}\,F\left(\frac12\,,\,\frac56\,;\, \frac{11}{6}\,;\, -\frac{\Omega_\Lambda}{\Omega_m}\,a^3 \right) \,+\, \frac{B(\bm{x})}{a^3} \,,
}
where $B(\bm{x})$ is the integration constant, and $F$ denotes Gauss's hypergeometric function. Integrating once again with respect to time, and using (\ref{second-integral}), we get
\eq{ \label{linear-E-general}
E = \frac{2\,A(\bm{x})}{5H_0^2\,\Omega_m} \, a\,F\left(\frac13\,,\,1\,;\, \frac{11}{6}\,;\,-\frac{\Omega_\Lambda}{\Omega_m}\,a^3\right) \,-\, \frac{2\,B(\bm{x})}{3H_0\,\sqrt{\Omega_m}\,a^{\frac32}}\,\sqrt{1 + \frac{\Omega_\Lambda}{\Omega_m}\,a^3} \,+\,C(\bm{x})\,,
}
where $C(\bm{x})$ is the integration constant. It represents a residual gauge freedom, which is not eliminated by choosing Lagrangian coordinates. This expression for the scalar mode $E$ matches the ones found in \cite{Silveira:1994, Vale:2001}, for the linear matter density perturbation in the presence of a cosmological constant.

As argued earlier, we choose initial conditions ensuring that no back-reaction is initially produced by the perturbations, at a time in the early stages of matter domination, when the scale factor is much smaller than its present value. In particular, as can be inferred from (\ref{linear-kinematical-1}), the kinematical back-reaction vanishes initially at leading order in the perturbations, $Q(a_\text{in}) = 0$, if we set $\dot{E}(a_\text{in},\bm{x}) = 0$. This allows us to fix the integration constant $B(\bm{x})$. Assuming that $a_\text{in} \ll 1$, we are then left with 
\eqa{ \label{linear-E}
&\dot{E} = \frac{A(\bm{x})}{H_0}\,f(a)\,, \\[5pt]
&E = \frac{A(\bm{x})}{H_0^2}\,g(a)\,+\, C(\bm{x})\,,
}
where the integration constant was redefined to correspond to the initial value of the perturbation, $C(\bm{x}) = E(a_\text{in}, \bm{x})$. Moreover, $f$ and $g$ denote the following functions of the scale factor, both of which vanish initially as long as $a_\text{in} \ll 1$,
\eqa{ \label{functions-f-g}
f(a) &= \frac{2}{5\sqrt{\Omega_m}\,a^{\frac12}}\left[ F\left(\frac12\,,\,\frac56\,;\, \frac{11}{6}\,;\, -\frac{\Omega_\Lambda}{\Omega_m}\,a^3 \right) - \left(\frac{a_\text{in}}{a}\right)^{\frac52}\right] \,, \\[5pt]
g(a) &= \frac{2}{5\,\Omega_m}\,\,a\left[ \,F\left(\frac13\,,\,1\,;\, \frac{11}{6}\,;\,-\frac{\Omega_\Lambda}{\Omega_m}\,a^3\right) + \frac23 \left( \frac{a_\text{in}}{a}\right)^{\frac52}\sqrt{1 + \frac{\Omega_\Lambda}{\Omega_m}\,a^3} \,-\, \frac{5a_\text{in}}{3a}\,\,\right]\,.
}

At late times, when the scale factor becomes much larger than its initial value, the terms in these functions involving $a_\text{in}$ become negligible. Far into the future, when the scale factor is much larger than its present value, $a \gg 1$, these functions behave asymptotically as
\eqa{ \label{functions-f-g-asymptotic}
f(a) &= \frac{1}{\sqrt{\Omega_\Lambda}\,a^2}\,+\, \order{\frac{1}{a^3}}\,, \\[7pt]
g(a) &= \frac{2\,\Gamma\left(\frac{11}{6}\right)\Gamma\left(\frac{2}{3}\right)}{5\,\Gamma\left(\frac{3}{2}\right)\,\Omega_m^{\frac23}\,\Omega_\Lambda^{\frac13}} \,-\, \frac{1}{2\,\Omega_\Lambda\,a^2} \,+\, \order{\frac{1}{a^3}}\,.
}

\subsection{Average spatial curvature} \label{section-perturbation-curvature}

The back-reaction terms in the average Friedmann equations include the average spatial curvature (or Ricci scalar) of the three-dimensional metric $g_{ij}$. Expanding the Ricci scalar up to the second order in the scalar modes of the perturbations, we find
\eqa{ \label{Ricci-second-order}
R &= -2\,\cov^2 \psi + 4\,\psi \cov^2 \psi + \frac32 \p_i \psi \,\p_i \psi + \cov^2 \psi \,\cov^2 E + \p_i\p_j \psi \,\p_i\p_j E + \p_i \psi \,\p_i\cov^2 E\\[3pt] 
&\hspace{30pt} - \frac14 \p_i \cov^2 E \,\p_i \cov^2 E + \frac14 \p_i\p_j\p_k E \,\p_i\p_j\p_k E\,.
}

The Ricci scalar is linear in the perturbation $\psi$, which is constant at first order, but ceases to be at higher orders. Hence, we cannot simply plug $\psi = A(\bm{x})$ in the linear term, since its second-order time-dependence should also be included. The mode $\psi$ is determined by the momentum constraint (\ref{Einstein-momentum}). At the second order in the scalar perturbations, this constraint implies
\eq{
\p_i \dot{\psi} - \frac12 \cov^2 \dot{E}_i = \frac14 \p_j \psi\,\p_i\p_j \dot{E} + \frac14 \p_i \psi\,\cov^2\dot{E} - \frac14 \p_j \cov^2 E \,\p_i\p_j \dot{E} + \frac14 \p_i\p_j\p_k E\,\p_j\p_k \dot{E}\,,
}
where $E_i$ represents the second-order scalar-induced vector modes. Integrating with respect to time, and using the first-order solutions (\ref{linear-psi}) and (\ref{linear-E}), we get
\eqa{
\p_i \psi - \frac12 \cov^2 E_i &= \p_i A + \frac14 \p_j A\,\p_i\p_j E + \frac14 \p_i A\,\cov^2E - \frac18 \p_j \cov^2 E \,\p_i\p_j E \\[3pt]
&\hspace{30pt} + \frac18 \p_i\p_j\p_k E\,\p_j\p_k E + A_i^{(2)}\,,
}
where $A(\bm{x})$ denotes the constant value of the mode $\psi$ at linear order, and $A_i^{(2)}(\bm{x})$ is the second-order integration constant. Taking the divergence of this equation to get rid of the vector modes, we find
\eqa{
\cov^2\psi &= \cov^2 A + \frac14 \cov^2A \,\cov^2E + \frac12 \p_iA\,\p_i\cov^2E + \frac14 \p_i\p_jA \,\p_i\p_j E + \frac14 \p_i\p_j A \,\p_i\p_j E \\[3pt]
&\hspace{30pt} -\frac18 \p_i \cov^2 E\,\p_i\cov^2 E + \frac18 \p_i\p_j\p_k E\,\p_i\p_j\p_k E + \p_i A_i^{(2)}\,.
}
Plugging this equation in (\ref{Ricci-second-order}), the spatial curvature can then be written in terms of its initial value, plus other terms which vary with time,
\eq{ \label{linear-curvature}
R = \bar{R} + \frac{1}{2H_0^2}\,\Big( \p_i\p_j A\,\p_i\p_j A + (\cov^2A)^2 \Big) \,g(a)\,,
}
where $\bar{R}$ represents the initial value of the Ricci scalar,
\eq{
\bar{R} = -2\,\cov^2 A + 4\,A \,\cov^2 A + \frac32 \p_i A \,\p_i A + \frac12 \cov^2 A \,\cov^2 C + \frac12 \p_i\p_j A \,\p_i\p_j C - 2\,\p_i A_i^{(2)}\,.
}

When computing the average spatial curvature over the volume of  chosen particles, we must keep in mind that the volume itself changes with time because of the spacetime expansion. To factor out this time dependence, we expand the average in terms of its initial value, taken over the initial volume occupied by the particles. Details about the calculation are provided in appendix \ref{appendix-average}. At first order in the perturbations, the volume of a small region of particles can be expanded as
\eq{
\sqrt{g^{(3)}} = 1 + \frac32 A + \frac12 \cov^2 E\,.
}
that is, in terms of its initial value $\sqrt{\bar{g}^{(3)}}$, 
\eq{ \label{linear-volume}
\sqrt{g^{(3)}} = \sqrt{\bar{g}^{(3)}}\left( 1 + \frac{1}{2H_0^2}\,\cov^2A\,\,g(a)\right)\,.
}
Plugging (\ref{linear-curvature}) and (\ref{linear-volume}) in equation (\ref{appendix-average-result}) from appendix \ref{appendix-average}, the average spatial curvature is then
\eq{ \label{linear-average-curvature}
\frac{\expval{R}}{a^2} = \frac{\expval{\bar{R}}_\text{in}}{\bar{a}^2} + \frac{1}{H_0^2}\left(\frac12\expval{\p_i\p_jA\,\p_i\p_j A} - \frac12\expval{(\cov^2 A)^2} + \frac13\,\expval{\cov^2 A}^2 \right) \frac{g(a)}{a^2}\,,
}
where the subscript ``in'' indicates that the average is taken over the volume initially occupied by the chosen particles. We will drop this subscript whenever taking the average over the initial volume is the same as over the actual volume at a given perturbative order. Moreover, $\bar{a}$ represents the effective scale factor associated with the volume of the chosen particles as it expands. We will use $a$ instead whenever it coincides with the background scale factor at a given perturbative order.

\subsection{Quadratic back-reaction from scalar modes}

We can now compute the back-reaction energy density and pressure generated at quadratic order by the scalar perturbations. Since $L_i^j = \p_i\p_j \dot{E}$ at linear order, the kinematical back-reaction (\ref{kinematical}) is then
\eq{ \label{linear-kinematical-1}
Q = \frac18 \expval{\p_i\p_j \dot{E}\,\p_i\p_j \dot{E}} - \frac18 \expval{(\cov^2 \dot{E})^2} + \frac{1}{12} \expval{\cov^2 \dot{E}}^2\,.
}
Using the solution obtained earlier for the perturbation (\ref{linear-E}), it becomes
\eq{ \label{linear-kinematical}
Q = \frac{1}{H_0^2}\left( \frac18 \expval{\p_i\p_j A\,\p_i\p_j A} - \frac18 \expval{(\cov^2 A)^2} + \frac{1}{12}\expval{\cov^2 A}^2 \right) f(a)^2\,.
}

This kinematical back-reaction and the average spatial curvature  are part of the average Friedmann equations. Among these contributions, only the initial value of the average spatial curvature behaves as constant spatial curvature, having an energy density $\rho_K$ and a pressure $p_K = -\frac13 \rho_K$ given by
\eq{ \label{constant-curvature}
\kappa\rho_K = -\frac{\expval{\bar{R}}_\text{in}}{2\bar{a}^2}\,.
}

The other terms do not behave as constant spatial curvature, having a different dependence on the scale factor, and thus form the back-reaction energy density and pressure. Plugging (\ref{linear-average-curvature}), (\ref{linear-kinematical}), and (\ref{constant-curvature}) in (\ref{br-density-pressure}), we find
\eqa{ \label{linear-back-reaction}
&\kappa \rho_\text{br} = -\frac{1}{4H_0^2}\left( \big\langle \p_i \big( \p_j A \,\p_i\p_j A\big)\big\rangle - \big\langle \p_i \big( \p_i A\,\cov^2 A\big)\big\rangle + \frac23 \expval{\cov^2 A}^2 \right) \left( \frac{g(a)}{a^2} - \frac12\,f(a)^2 \right) \,, \\[5pt]
&\kappa p_\text{br} = \frac{1}{4H_0^2}\left( \big\langle \p_i \big( \p_j A \,\p_i\p_j A\big)\big\rangle - \big\langle \p_i \big( \p_i A\,\cov^2 A\big)\big\rangle + \frac23 \expval{\cov^2 A}^2 \right) \left( \frac{g(a)}{3a^2} + \frac12\,f(a)^2 \right) \,.
}

\subsection{Dark energy equation of state}

As argued in the introduction of this paper, combining the back-reaction of perturbations with the cosmological constant produces an effective dark energy which can evolve with time. Plugging (\ref{linear-back-reaction}) in (\ref{DE-equation-state}), its equation of state at quadratic order in the scalar modes of the perturbations is then
\eqa{ \label{linear-DE-equation}
w_\text{DE} &= -1 - \frac{1}{18H_0^4\,\Omega_\Lambda}\left( \big\langle \p_i \big( \p_j A \,\p_i\p_j A\big)\big\rangle - \big\langle \p_i \big( \p_i A\,\cov^2 A\big)\big\rangle + \frac23 \expval{\cov^2 A}^2 \right) \left( \frac{g(a)}{a^2} - \frac32\,f(a)^2 \right)\,.
}

This important result is what we set out to compute in this section. It coincides with the equation of state discussed in the introduction (\ref{linear-intro}). The back-reaction terms correspond to boundary terms, which become negligible when the average is taken globally, over all matter particles. Scalar perturbations can generate significant back-reaction even in the case of a global average only beyond quadratic order, at which point scalar-induced vector and tensor modes must also be taken into account.

The equation of state that we obtained is valid at all times. It can be expanded around present time, when the scale factor is close to $a=1$, to match the Chevallier-Polarski-Linder parametrization of dark energy. Using (\ref{equation-E}) and (\ref{linear-E}), the derivatives of the functions $f(a)$ and $g(a)$ with respect to the scale factor are
\eqa{
&\dv{f}{a} = -\frac{3}{a}\,f(a) + \frac{H_0}{a^3\,H}\,, \\[5pt]
&\dv{g}{a} = \frac{H_0}{a\,H}\,f(a)\,.
}
Linearizing these functions around present time, we find
\eqa{ \label{functions-linearized}
f(a) &= f_0 + (3f_0 - 1)(1-a) \,+\, \order{(1-a)^2}\,, \\[5pt]
g(a) &= g_0 - f_0\,(1-a) \,+\, \order{(1-a)^2}\,,
}
where $f_0=f(1)$ and $g_0=g(1)$ denote the values of the functions at present time. Since initial conditions are fixed in the early stages of matter domination, when the scale factor is much smaller than its present value, the terms involving the initial value of the scale factor $a_\text{in}$ in (\ref{functions-f-g}) can now be safely neglected,
\eqa{ \label{functions-f-g-present}
&f_0 = \frac{2}{5\sqrt{\Omega_m}}\, F\left(\frac12\,,\,\frac56\,;\, \frac{11}{6}\,;\, -\frac{\Omega_\Lambda}{\Omega_m} \right) + \order{a_\text{in}^{\frac52}}\,, \\[5pt]
&g_0 = \frac{2}{5\,\Omega_m} \,F\left(\frac13\,,\,1\,;\, \frac{11}{6}\,;\,-\frac{\Omega_\Lambda}{\Omega_m}\right) + \order{a_\text{in}}\,.
}

For example, when $\Omega_m = 0.3$ and $\Omega_\Lambda = 0.7$, we find $f_0 = 0.53261$ and $g_0 = 1.03864$. Using these linear expansions, the parameters involved in the equation of state of dark energy around present time (\ref{CPL parametrization}) are then
\eqa{ \label{linear-parameters}
\delta w_0 &= -\frac{1}{18H_0^4\,\Omega_\Lambda}\left( \big\langle \p_i \big( \p_j A \,\p_i\p_j A\big)\big\rangle - \big\langle \p_i \big( \p_i A\,\cov^2 A\big)\big\rangle + \frac23 \expval{\cov^2 A}^2 \right) \left( g_0 - \frac32\,f_0^2 \right)\,, \\[5pt]
\delta w_1 &= -\frac{1}{9H_0^4\,\Omega_\Lambda}\left( \big\langle \p_i \big( \p_j A \,\p_i\p_j A\big)\big\rangle - \big\langle \p_i \big( \p_i A\,\cov^2 A\big)\big\rangle + \frac23 \expval{\cov^2 A}^2 \right) \left( g_0 + f_0 - \frac92\,f_0^2 \right) \,.
}

\subsection{Linear tensor modes}

Let us now consider the tensor modes. Since $L_i^j = \dot{H}_{ij}$ at linear order, the kinematical back-reaction (\ref{kinematical}) generated at quadratic order by the tensor modes is
\eq{ \label{tensor-kinematical}
Q = \frac18 \expval{\dot{H}_{ij} \dot{H}_{ij}}\,.
}
The kinematical back-reaction vanishes initially if we set $\dot{H}_{ij}(a_\text{in}, \bm{x}) = 0$. At linear order, the tensor modes satisfy the following wave equation,  
\eq{
\ddot{H}_{ij} + 3H\,\dot{H}_{ij} - \frac{1}{a^2}\,\cov^2H_{ij}  = 0\,.
}

This wave equation admits well-known solutions in certain limits, for example when the spacetime expansion is dominated either by the cold matter or the cosmological constant. However, we have not (yet) found in compact form its full solution, valid at all times. For this reason, we choose to solve the wave equation with a gradient expansion, by treating spatial derivatives perturbatively. 

The first term of the gradient expansion corresponds to the long-wavelength limit of the perturbations, when their spatial derivatives can be neglected. In that limit, the wave equation becomes $\ddot{H}_{ij} + 3H\,\dot{H}_{ij} = 0$. The only solution also satisfying the initial condition corresponds to a constant, $H_{ij} = A_{ij}(\bm{x})$. 

The second term in the gradient expansion can be obtained by plugging this long-wavelength limit back into the original wave equation, wherever spatial derivatives appear. The wave equation then becomes
\eq{ \label{equation-tensor}
\ddot{H}_{ij} + 3H\,\dot{H}_{ij} - \frac{1}{a^2}\,\cov^2A_{ij}  = 0\,.
}

This is formally equivalent to the equation (\ref{equation-E}) satisfied by the scalar mode $E$. Using the integrals from appendix \ref{appendix-integrals}, we find
\eqa{
&\dot{H}_{ij} = \frac{\cov^2 A_{ij}}{H_0}\,f(a)\,, \\[3pt]
&H_{ij} = A_{ij} + \frac{\cov^2 A_{ij}}{H_0^2}\,g(a) \,,
}
where $f(a)$ and $g(a)$ are the same functions defined earlier (\ref{functions-f-g}). Hence, the two first terms of the gradient expansion of the linear tensor modes are formally similar to the linear scalar modes, involving the same functions of the scale factor. Plugging this solution into (\ref{tensor-kinematical}), the kinematical back-reaction from the linear tensor modes is then
\eq{
Q = \frac{1}{8H_0^2} \expval{\cov^2 A_{ij} \, \cov^2 A_{ij}}\,.
}

The back-reaction from the two first terms of the gradient expansion does not reduce to a boundary term and can thus remain significant even when the average is taken globally, over all matter particles. As argued in the introduction, the same should hold for the linear tensor modes in general, when their full expression is taken into account.

\section{Gradient expansion} \label{section-gradient}

\subsection{Relevance of a gradient expansion}

As shown in the previous computation, the back-reaction terms generated by the scalar modes of the perturbations reduce to surface terms at quadratic order, thus making them negligible when the average is taken globally. Although their contribution at higher orders can be significant even for a global average, we would then need to compute the second-order vector and tensor modes that they induce. However, we cannot solve fully the wave equation satisfied by the linear tensor modes, let alone the same wave equation in the presence of a source term. 

For these reasons, a different approach is needed. In this section, we will no longer treat the inhomogeneities as small perturbations affecting a homogeneous background. Instead, we will solve the full Einstein equations with a gradient expansion, by treating spatial derivatives perturbatively, rather than the inhomogeneities themselves. A gradient expansion was previously used in \cite{Afshordi:2000, Brandenberger:2018}, when computing the back-reaction of inhomogeneities on the average expansion of a spacetime containing scalar fields.

As for the linear tensor modes considered earlier, the back-reaction terms generated at the second order in the gradient expansion do not reduce to surface terms. They involve the same functions of the scale factor as the linear scalar modes. In fact, if the inhomogeneities are once again treated as small perturbations, the back-reaction terms reduce to the ones generated at quadratic order by the scalar modes. This is not surprising, since the equations satisfied by the scalar modes do not contain any spatial derivative. Their long-wavelength limit trivially coincides with their full solution.

Hence, the gradient expansion allows us to compute some of the back-reaction from the scalar modes not just at quadratic order in their amplitude, but at all orders. It also allows us to compute the back-reaction of the corresponding scalar-induced vector and tensor modes. Finally, it would be quite straightforward to extend our computation beyond the second order in the gradient expansion, since the equations encountered at higher orders can still be solved in terms of hypergeometric functions of the scale factor. The same cannot be said about the standard perturbative approach.

\subsection{Second-order gradient expansion}

In Lagrangian coordinates, the spatial derivatives involved in the  Einstein equations are contained either in covariant derivatives or in the Ricci tensor of the three-dimensional metric $g_{ij}$. In the long-wavelength limit, when neglecting all spatial derivatives, the evolution equation (\ref{Einstein-evolution}) becomes
\eq{
\dot{L}_i^j + 3H L_i^j + \frac12 L L_i^j + \frac18\,\delta_i^j\Big(L_k^l L_l^k - L^2\Big) = 0\,.
}

We can find non-trivial solutions to this equation, which will be discussed in more details elsewhere. However, for the purposes of this paper, we will focus on the simple solution when $L_i^j = g^{jk}\,\dot{g}_{ik} = 0$, or equivalently when the spatial components of the metric do not change with time, $g_{ij} = \bar{g}_{ij}(\bm{x})$. In particular, this constant value includes the linear scalar mode $\psi = A(\bm{x})$, as well as the long-wavelength limit of the linear tensor modes, $H_{ij} = A_{ij}(\bm{x})$. Hence, this solution will allow us to compute the back-reaction of these linear modes not just at quadratic order, but at all orders in their amplitude.

At the second order in the gradient expansion, the metric can be written as its constant long-wavelength limit, plus a small perturbation which can vary with time,
\eq{
g_{ij} = \bar{g}_{ij}(\bm{x}) + h_{ij}(a, \bm{x})\,. 
}

In other words, the perturbations are added to an inhomogeneous background, whose spatial curvature is not necessarily constant. The background inhomogeneities do not themselves produce back-reaction, since their contribution to the average Friedmann equations behaves like constant spatial curvature. Only higher order terms in the gradient expansion produce actual back-reaction and can thus generate evolving dark energy. Moreover, the distinction between the background and the perturbations is defined up to a constant. The background can thus be made to coincide initially with the spatial components of the metric,
\eq{ \label{gradient-initial}
\bar{g}_{ij}(\bm{x}) = g_{ij}(a_\text{in}, \bm{x})\,,
}
which implies that the perturbations vanish initially, $h_{ij}(a_\text{in}, \bm{x}) = 0$. Furthermore, as can be inferred from (\ref{gradient-kinematical}), the kinematical back-reaction vanishes initially if we set $\dot{h}_{ij}(a_\text{in}, \bm{x}) = 0$. Despite these initial conditions, the perturbations are not identically zero, being driven away from zero by the background inhomogeneities.

In this section, indices of the perturbations will be raised and lowered with the background metric, for instance $h_i^j = \bar{g}^{jk} \,h_{ik}$. At first order in the perturbations, we find $L_i^j = \dot{h}_i^j$. Neglecting higher-order terms in the perturbations and terms involving their spatial derivatives, the evolution equation (\ref{Einstein-evolution}) becomes
\eq{
\ddot{h}_{ij} + 3H\,\dot{h}_{ij} + \frac{1}{a^2}\left(2\bar{R}_{ij} - \frac12 \bar{R}\,\bar{g}_{ij}\right) = 0\,,
} 
where $\bar{R}_{ij}$ represents the Ricci tensor of the three-dimensional background metric $\bar{g}_{ij}$, and $\bar{R} = \bar{g}^{ij}\,\bar{R}_{ij}$ is its Ricci scalar. This equation is formally equivalent to the one satisfied by the linear scalar mode $E$, as well as by the second term of the gradient expansion of the linear tensor modes, namely equations (\ref{equation-E}) and (\ref{equation-tensor}). Integrating with respect to time, using the integrals from appendix \ref{appendix-integrals}, and applying the initial conditions discussed above, we find
\eqa{ \label{gradient-solutions}
&\dot{h}_{ij} = -\frac{1}{H_0}\left(2\bar{R}_{ij} - \frac12 \bar{R}\,\bar{g}_{ij}\right) f(a) \,, \\[5pt]
&h_{ij} = -\frac{1}{H_0^2}\left(2\bar{R}_{ij} - \frac12 \bar{R}\,\bar{g}_{ij}\right) \,g(a) \,,
}
where $f(a)$ and $g(a)$ are the same functions defined earlier (\ref{functions-f-g}).

\subsection{Average spatial curvature} \label{section-gradient-curvature}

The back-reaction terms in the average Friedmann equations include the average spatial curvature (or Ricci scalar) of the three-dimensional metric $g_{ij}$. At the second order in the gradient expansion, it suffices to expand the Ricci scalar at the first order in the perturbations, since higher-order terms involve more than four spatial derivatives. According to the well-known formula,
\eq{
R = \bar{R} + \bar{\cov}_i\bar{\cov}^j h_j^i - \bar{\cov}_i \bar{\cov}^i h - \bar{R}_i^j \,h_j^i \,,
}
where $\bar{\cov}_i$ represents the covariant derivative for the background metric $\bar{g}_{ij}$. Using the solution obtained earlier for the perturbations (\ref{gradient-solutions}), and the fact that $\bar{\cov}_j \bar{R}_i^j = \frac12 \p_i\bar{R}$, we find
\eq{ \label{gradient-curvature}
R = \bar{R} \,+ \frac{1}{H_0^2}\left( 2\bar{R}_i^j\,\bar{R}_j^i - \frac12 \bar{R}^2\right) g(a)\,.
}

When computing the average spatial curvature, we also want to factor out the time-dependence of the volume of particles over which the average is taken. At first order in the perturbations, the volume of a small region of particles can be expanded as
\eq{ \label{gradient-volume}
\sqrt{g^{(3)}} = \sqrt{\bar{g}^{(3)}}\left(1 + \frac12 h \right)
= \sqrt{\bar{g}^{(3)}} \left( 1 - \frac{1}{4H_0^2} \,\bar{R}\,g(a) \right) \,,
}
where $\sqrt{\bar{g}^{(3)}}$ represents its initial value, when the volume has the same metric as the background. Plugging (\ref{gradient-curvature}) and (\ref{gradient-volume}) in equation (\ref{appendix-average-result}) from appendix \ref{appendix-average}, the average spatial curvature is then
\eq{ \label{gradient-average-curvature}
\frac{\expval{R}}{a^2} = \frac{\expval{\bar{R}}_\text{in}}{\bar{a}^2} + \frac{2}{H_0^2}\left( \expval{\bar{R}_i^j \,\bar{R}_j^i} - \frac38 \expval{\bar{R}^2} + \frac{1}{24} \expval{\bar{R}}^2\right)\frac{g(a)}{a^2}\,,
}
where the subscript ``in'' indicates that the average is taken over the volume initially occupied by the chosen particles. Moreover, $\bar{a}$ represents the effective scale factor defined by their volume at a given time.

\subsection{Dark energy equation of state}

We can now compute the back-reaction energy density and pressure generated by the inhomogeneities at the second order in the gradient expansion. Since $L_i^j = \dot{h}_i^j$ at first order, the kinematical back-reaction (\ref{kinematical}) is
\eq{ \label{gradient-kinematical-1}
Q = \frac18 \expval{ \dot{h}_i^j \,\dot{h}_j^i } - \frac18 \expval{\dot{h}^2} + \frac{1}{12}\expval{\dot{h}}^2\,.
}
Using the solution obtained earlier for the perturbation (\ref{gradient-solutions}), it becomes
\eq{ \label{gradient-kinematical}
Q = \frac{1}{2H_0^2} \left( \expval{\bar{R}_i^j \,\bar{R}_j^i} - \frac38 \expval{\bar{R}^2} + \frac{1}{24} \expval{\bar{R}}^2\right) f(a)^2\,.
}

Once again, only the initial value of the average spatial curvature behaves as constant spatial curvature, having an energy density $\rho_K$ and a pressure $p_K = -\frac13 \rho_K$ given as before by (\ref{constant-curvature}). The other terms contribute instead to the back-reaction energy density and pressure. Plugging (\ref{constant-curvature}), (\ref{gradient-average-curvature}), and (\ref{gradient-kinematical}) in (\ref{br-density-pressure}), we find
\eqa{ \label{gradient-back-reaction}
&\kappa \rho_\text{br} = -\frac{1}{H_0^2} \left(\expval{\bar{R}_i^j \,\bar{R}_j^i} - \frac38 \expval{\bar{R}^2} + \frac{1}{24} \expval{\bar{R}}^2 \right)\left( \frac{g(a)}{a^2} - \frac12 \,f(a)^2\right) \\[5pt]
&\kappa p_\text{br} = \frac{1}{H_0^2} \left(\expval{\bar{R}_i^j \,\bar{R}_j^i} - \frac38 \expval{\bar{R}^2} + \frac{1}{24} \expval{\bar{R}}^2 \right)\left( \frac{g(a)}{3a^2} + \frac12\,f(a)^2\right) \,.
}

Combining these equations with (\ref{DE-equation-state}), the equation of state of dark energy generated by the inhomogeneities at the second order in the gradient expansion is then
\eq{ \label{gradient-DE-equation}
w_\text{DE} = -1 \,-\, \frac{2}{9H_0^4\,\Omega_\Lambda}\left(\expval{\bar{R}_i^j\,\bar{R}_j^i} - \frac38 \expval{\bar{R}^2} + \frac{1}{24} \expval{\bar{R}}^2 \right) \left( \frac{g(a)}{a^2} - \frac32\,f(a)^2\right) \,.
}

This is perhaps the most important result of this paper. If the inhomogeneities are once again treated as small perturbations affecting a homogeneous background, with $\bar{g}_{ij} = \delta_{ij} + \delta_{ij}\,A(\bm{x})$, the previous equations can be expanded at the second order in the scalar mode $A(\bm{x})$. It then coincides with the equation of state obtained following the standard perturbative approach (\ref{linear-DE-equation}), thus confirming that a gradient expansion was an appropriate way of generalizing our perturbative calculation. 

Here, the back-reaction does not reduce to surface terms and remains significant even for a global average. In particular, it comprises the effects from the linear scalar mode $\psi = A(\bm{x})$ at all orders in its amplitude, as well as those from the long-wavelength limit of the linear tensor modes $H_{ij} = A_{ij}(\bm{x})$, and from their interactions.

Linearizing this equation around present time with (\ref{functions-linearized}), the Chevallier-Polarski-Linder parametrization of the equation of state of dark energy (\ref{CPL parametrization}) then involves the following parameters,
\eqa{ \label{gradient-parameters}
&\delta w_0 = -\frac{2}{9H_0^4\,\Omega_\Lambda}\left(\expval{\bar{R}_i^j\,\bar{R}_j^i} - \frac38 \expval{\bar{R}^2} + \frac{1}{24} \expval{\bar{R}}^2 \right)\left(g_0 - \frac32 f_0^2\right) \,, \\[5pt]
&\delta w_1 = -\frac{4}{9H_0^4\,\Omega_\Lambda}\left(\expval{\bar{R}_i^j\,\bar{R}_j^i} - \frac38 \expval{\bar{R}^2} + \frac{1}{24} \expval{\bar{R}}^2 \right)\left(g_0 + f_0 - \frac92 f_0^2 \right) \,.
}
where $f_0$ and $g_0$ denote the values of the functions at the present time (\ref{functions-f-g-present}). They depend exclusively on the density parameters of cold matter and the cosmological constant. For example, if $\Omega_m = 0.3$ and $\Omega_\Lambda = 0.7$, the parameters are then
\eqa{
&\delta w_0 = -\frac{0.1946}{H_0^4}\left(\expval{\bar{R}_i^j\,\bar{R}_j^i} - \frac38 \expval{\bar{R}^2} + \frac{1}{24} \expval{\bar{R}}^2\right) \,, \\[5pt]
&\delta w_1 = -\frac{0.1871}{H_0^4}\left(\expval{\bar{R}_i^j\,\bar{R}_j^i} - \frac38 \expval{\bar{R}^2} + \frac{1}{24} \expval{\bar{R}}^2 \right) \,.
}

These parameters were discussed in the introduction. Similarly, when the scale factor becomes much larger than its present value, $a \gg 1$, the equation of state behaves asymptotically as
\eq{
w_\text{DE} = -1 - \frac{4\,\Gamma\left(\frac{11}{6}\right)\Gamma\left(\frac{2}{3}\right)}{45\,H_0^4\,\Gamma\left(\frac{3}{2}\right)\,\Omega_m^{\frac23}\,\Omega_\Lambda^{\frac43}} \left(\expval{\bar{R}_i^j\,\bar{R}_j^i} - \frac38 \expval{\bar{R}^2} + \frac{1}{24} \expval{\bar{R}}^2 \right) \frac{1}{a^2} \,+\, \order{\frac{1}{a^4}} \,,
}
where we used (\ref{functions-f-g-asymptotic}). As expected, the time dependence induced by the inhomogeneities disappears far into the future, when the spacetime becomes dominated by the cosmological constant.

\begin{appendices}
\section{Two Integrals} \label{appendix-integrals}

In this appendix, we compute two integrals which came up when solving the differential equation satisfied by metric perturbations. The results are
\eq{ \label{first-integral}
\int \dd{t} a \,=\, \frac{2a^{\frac52}}{5H_0\sqrt{\Omega_m}}\,F\left(\frac12\,,\,\frac56\,;\, \frac{11}{6}\,;\, -\frac{\Omega_\Lambda}{\Omega_m}\,a^3 \right)\,,
} \vspace{-20pt}
\eq{ \label{second-integral}
\int \frac{\dd{t}}{a^{\frac12}}\,F\left(\frac12\,,\,\frac56\,;\, \frac{11}{6}\,;\, -\frac{\Omega_\Lambda}{\Omega_m}\,a^3 \right) \,=\, 
\frac{a}{H_0\sqrt{\Omega_m}}\,F\left(\frac13\,,\,1\,;\, \frac{11}{6}\,;\,-\frac{\Omega_\Lambda}{\Omega_m}\,a^3\right)\,,
}
where $F$ represents Gauss's hypergeometric function. For general coefficients, it is defined by the series
\eq{ \label{hypergeometric}
F(a\,,\,b\,;\,c\,;\,x) = \sum_{n\geq0} \frac{(a)_n\,(b)_n}{n!\,(c)_n}\,x^n\,,
}
where $(x)_n = x(x+1)(x+2)\cdots(x+n-1)$ denotes the rising factorial, also known as the Pocchammer symbol.

\subsection{First integral}
The first integral is
\eq{
\int \dd{t} a = \int \frac{\dd{a}}{H}\,.
}
For a spacetime containing cold matter and a cosmological constant, the background expansion rate is given by equation (\ref{background-H}), which can be rewritten as
\eq{
H = \frac{H_0\sqrt{\Omega_m}}{a^{\frac32}} \,\sqrt{1 + \frac{\Omega_\Lambda}{\Omega_m}\,a^3}\,.
}
Expanding the square root using Newton's binomial formula, we find
\eq{ \label{H-inverse-series}
\frac{1}{H} = \frac{a^{\frac32}}{H_0\sqrt{\Omega_m}}\,\sum_{n\geq0} \frac{\left(\frac12\right)_n}{n!} \,\left(-\frac{\Omega_\Lambda}{\Omega_m}\,a^3\right)^n\,,
}
Integrating with respect to the scale factor, we find
\eqa{
\int \frac{\dd{a}}{H} &= \frac{2a^{\frac52}}{5H_0\sqrt{\Omega_m}}\,\sum_{n\geq0} \frac{\left(\frac12\right)_n}{n!} \frac{1}{1 + \frac65 n} \,\left(-\frac{\Omega_\Lambda}{\Omega_m}\,a^3\right)^n \\[5pt]
&= \frac{2a^{\frac52}}{5H_0\sqrt{\Omega_m}}\,\sum_{n\geq0} \frac{\left(\frac12\right)_n \left(\frac56\right)_n}{n!\,\left(\frac{11}{6}\right)_n} \,\left(-\frac{\Omega_\Lambda}{\Omega_m}\,a^3\right)^n \\[5pt]
&= \frac{2a^{\frac52}}{5H_0\sqrt{\Omega_m}}\,F\left(\frac12\,,\,\frac56\,;\, \frac{11}{6}\,;\, -\frac{\Omega_\Lambda}{\Omega_m}\,a^3 \right)\,,
}
where $F$ represents Gauss's hypergeometric function, defined in (\ref{hypergeometric}).

\subsection{Second integral}

Similarly, the second integral that we need to compute is
\eqa{
\int \frac{\dd{t}}{a^{\frac12}}\,F\left(\frac12\,,\,\frac56\,;\, \frac{11}{6}\,;\, -\frac{\Omega_\Lambda}{\Omega_m}\,a^3 \right) \,=\, \int \frac{\dd{a}}{a^{\frac32}\,H}\,F\left(\frac12\,,\,\frac56\,;\, \frac{11}{6}\,;\, -\frac{\Omega_\Lambda}{\Omega_m}\,a^3 \right)\,.
}
Expanding once again the background expansion rate and the hypergeometric function in series, using (\ref{hypergeometric}) and (\ref{H-inverse-series}), we find
\eq{
\frac{1}{a^{\frac32}\,H}\,F\left(\frac12\,,\,\frac56\,;\, \frac{11}{6}\,;\, -\frac{\Omega_\Lambda}{\Omega_m}\,a^3 \right) = \frac{1}{H_0\sqrt{\Omega_m}}\sum_{m\geq0} \sum_{n\geq0} \frac{\left(\frac12\right)_m\left(\frac12\right)_n \left(\frac56\right)_n}{m!\,n!\,\left(\frac{11}{6}\right)_n} \left(-\frac{\Omega_\Lambda}{\Omega_m}\,a^3\right)^{m\,+\,n}\,.
}
Defining a new summation index $k=m+n$, we can rewrite the double sum as
\eqa{
\frac{1}{a^{\frac32}\,H}\,F\left(\frac12\,,\,\frac56\,;\, \frac{11}{6}\,;\, -\frac{\Omega_\Lambda}{\Omega_m}\,a^3 \right) &= \frac{1}{H_0\sqrt{\Omega_m}}\sum_{k\geq0}\sum_{n=0}^k \frac{\left(\frac12\right)_{k-n}\left(\frac12\right)_n \left(\frac56\right)_n}{(k-n)!\,n!\,\left(\frac{11}{6}\right)_n} \left(-\frac{\Omega_\Lambda}{\Omega_m}\,a^3\right)^k \\[5pt]
&= \frac{1}{H_0\sqrt{\Omega_m}}\sum_{k\geq0} \frac{\left(\frac43\right)_k}{\left(\frac{11}{6}\right)_k} \left(-\frac{\Omega_\Lambda}{\Omega_m}\,a^3\right)^k\,,
}
where the finite sum over $n$ was computed using  Pfaff's summation formula. Integrating with respect to the scale factor, we get
\eqa{
\int \frac{\dd{a}}{a^{\frac32}\,H}\,F\left(\frac12\,,\,\frac56\,;\, \frac{11}{6}\,;\, -\frac{\Omega_\Lambda}{\Omega_m}\,a^3 \right) \,&=\,  \frac{a}{H_0\sqrt{\Omega_m}}\,\sum_{k\geq0} \frac{\left(\frac43\right)_k}{\left(\frac{11}{6}\right)_k} \frac{1}{1 + 3k} \left(-\frac{\Omega_\Lambda}{\Omega_m}\,a^3\right)^k \\[5pt] 
&=\, \frac{a}{H_0\sqrt{\Omega_m}}\,\sum_{k\geq0} \frac{\left(\frac13\right)_k}{\left(\frac{11}{6}\right)_k} \left(-\frac{\Omega_\Lambda}{\Omega_m}\,a^3\right)^k \\[5pt]
&=\, \frac{a}{H_0\sqrt{\Omega_m}}\,F\left(\frac13\,,\,1\,;\, \frac{11}{6}\,;\,-\frac{\Omega_\Lambda}{\Omega_m}\,a^3\right)\,.
}

\section{Average spatial curvature} \label{appendix-average}

In this appendix, we provide more details regarding the computation of the average spatial curvature in sections \ref{section-perturbation-curvature} and \ref{section-gradient-curvature}. In both cases, the Ricci scalar of the three-dimensional metric $g_{ij}$ can be expanded as
\eq{
R = \bar{R} + \delta R\,,
}
where $\bar{R}$ represents its initial value, when the scale factor is equal to $a_\text{in}$. Moreover, $\delta R$ contains the time-dependence of the spatial curvature, which can be treated as a small perturbation for the cases considered in this paper. Similarly, the volume occupied by a small region of particules can also be expanded as
\eq{ \label{appendix-volume}
\sqrt{g^{(3)}} = \sqrt{\bar{g}^{(3)}}\left(1 + \delta g^{(3)}\right)\,,
}  
where $\sqrt{\bar{g}^{(3)}}$ represents its initial value. Plugging these two expansions in (\ref{definition-average}), the average spatial curvature at a given time is then
\eq{
\expval{R} = \frac{\expval{\bar{R} + \delta R + \delta g^{(3)}\,\bar{R} + \delta g^{(3)}\,\delta R}_\text{in}}{1 + \expval{\delta g^{(3)}}_\text{in}}\,.
}
Here, the subscript ``in'' indicates that the average is taken over the volume initially occupied by the chosen particles,
\eq{
\expval{F}_\text{in} = \frac{1}{V_\text{in}}\int \dd[3]{x} \sqrt{\bar{g}^{(3)}}\,F(t,\bm{x})\,,
}
where $V_\text{in}$ denotes the initial volume of the particles. Expanding up to the first order in the perturbations, we find
\eq{
\expval{R} = \expval{\bar{R}}_\text{in} + \expval{\delta R} + \expval{\delta g^{(3)}\,\bar{R}} - \expval{\delta g^{(3)}}\,\expval{\bar{R}}\,.
}

In the back-reaction energy density and pressure, the average spatial curvature is divided by the square of the background scale factor. According to \cite{Buchert:2000}, it should be expressed rather in terms of the effective scale factor $\bar{a}$, defined by the volume of the chosen particles. Combining (\ref{appendix-volume}) with (\ref{volume}) and (\ref{effective-a}), the background and effective scale factors are related to each other by
\eq{
\frac{V}{V_\text{in}} = \left(\frac{\bar{a}}{a_\text{in}}\right)^3 = \left(\frac{a}{a_\text{in}}\right)^3 \left(1 + \expval{\delta g^{(3)}}_\text{in} \right)\,,
}
that is, at first order in the perturbations,
\eq{
\bar{a} = a\left( 1 + \frac13 \,\expval{\delta g^{(3)}}\right)\,.
}
The back-reaction induced by the average spatial curvature is then
\eq{ \label{appendix-average-result}
\frac{\expval{R}}{a^2} = \frac{1}{\bar{a}^2}\left( \expval{\bar{R}}_\text{in} + \expval{\delta R} + \expval{\delta g^{(3)}\,\bar{R}} - \frac13\, \expval{\delta g^{(3)}}\,\expval{\bar{R}} \right) \,.
}

\end{appendices}

\section*{Acknowledgments}

The work done for this paper was supported in part by the Fonds de recherche du Qu\'ebec. I have also benefited greatly from discussions with my PhD advisor Robert Brandenberger (McGill University). The figures were generated using R. No artificial intelligence tools were used at any stage of this work.

\bibliographystyle{mybibstyle}
\bibliography{references}

\end{document}